\documentclass[groupedaddress,superscriptaddress,amsmath,amssymb,aps,prl,
twocolumn,10pt,floatfix,longbibliography]{revtex4-2}

\usepackage{svg} 
\usepackage{graphicx}
\usepackage{dcolumn}
\usepackage{bm}
\usepackage[bookmarks=true,colorlinks]{hyperref}

\usepackage{pdfpages}
\usepackage{etoolbox}
\usepackage{siunitx}
\makeatletter
\patchcmd{\@outputpage@head}{\@ifx{\LS@rot\@undefined}{}{\LS@rot}}{}{}{}
\makeatother

\graphicspath{ {./figs/} }

\usepackage[normalem]{ulem}

\begin{document}

\title{Dynamic Correlations of Frustrated Quantum Spins from High-Temperature Expansion}

\author{Ruben Burkard}
\affiliation{Institut für Theoretische Physik, Universit\"at T\"ubingen, Auf der Morgenstelle 14, 72076 T\"ubingen, Germany}

\author{Benedikt Schneider}
\affiliation{Department of Physics and Arnold Sommerfeld Center for Theoretical
Physics, Ludwig-Maximilians-Universit\"at M\"unchen, Theresienstr.~37,
80333 Munich, Germany}
\affiliation{Munich Center for Quantum Science and Technology (MCQST), 80799 Munich, Germany}

\author{Bj\"orn Sbierski}
\affiliation{Institut für Theoretische Physik, Universit\"at T\"ubingen, Auf der Morgenstelle 14, 72076 T\"ubingen, Germany}

\date{\today}

\begin{abstract}
For quantum spin systems in equilibrium, the dynamic structure factor (DSF) is among the most feature-packed experimental observables. However, from a theory perspective it is often hard to simulate in an unbiased and accurate way, especially for frustrated and high-dimensional models at intermediate temperature.
To address this challenge, we compute the DSF from a dynamic extension of the high-temperature expansion to frequency moments. 
We focus on nearest-neighbor Heisenberg models with spin-lengths $S=1/2$ and $1$. We provide comprehensive benchmarks and consider a variety of frustrated two- and three-dimensional antiferromagnets as applications. In particular we shed new light on the anomalous intermediate temperature regime of the $S=1/2$ triangular lattice model and reproduce the DSF measured recently for the $S=1$ pyrochlore material $\mathrm{NaCaNi}_2\mathrm{F}_7$. An open-source numerical implementation for arbitrary lattice geometries is also provided.
\end{abstract}

\maketitle
\emph{Introduction}.—Localized quantum spins with frustrated interactions are ubiquitous in modern condensed matter experiments ranging from solid-state Mott insulators \cite{broholm_quantum_2020} to atom-based analog quantum simulators \cite{browaeysManyBodyPhysics2020}. A major goal is the realization of exotic correlated and entangled low-energy states like quantum spin liquids (QSL) \cite{savaryQuantumSpin2016}. 
The dynamical structure factor (DSF) defined as spatial and temporal Fourier transform of the equilibrium two-point correlator $\left\langle S_{i}^{z}(t)S_{i^\prime}^{z}\right\rangle$
is routinely measured via inelastic neutron scattering (INS) \cite{khatua_experimental_2023} in the solid-state context. For cold-atom setups the DSF is accessible via Raman spectroscopy  \cite{prichard_observation_2025}. The DSF probes collective spin dynamics and contains rich information on (dipolar) excitations, quasiparticles (or the absence thereof) and even entanglement structure \cite{hauke_measuring_2016, laurell_witnessing_2025}.
From a theory perspective, however, a quantitative and unbiased calculation of the DSF for generic and possibly gapless frustrated quantum spin-$S$ models at low temperature $T$ is often challenging. With exact-diagonalization and its derivatives (e.g.~finite-$T$ Lanczos method \cite{prelovsek_dynamical_2021}) limited to small systems and quantum Monte Carlo hampered by the sign problem, tensor-network methods like density-matrix renormalization group (DMRG) are state-of-the-art. However, besides in one-dimension (1D) \cite{kish_high-temperature_2024}, current simulations are only feasible for $T=0$, small $S$ and are severely affected by finite-size effects and entanglement \cite{drescher_dynamical_2023,sherman_spectral_2023}. Diagrammatic approaches that provide spin correlations in imaginary (Matsubara) frequency like pseudo-fermionic functional RG \cite{reuther_j1_2010,niggemann_frustrated_2021,muller_pseudo-fermion_2024} or bold-line diagrammatic Monte Carlo \cite{kulagin_bold_2013} are successful for static correlations down to intermediate $J/T$ with $J$ the spin-spin interaction. However analytic continuation of this approximate Matsubara correlator to the real-frequency DSF is notoriously unstable. A popular bypass are semiclassical approximations of spin dynamics \cite{zhang_dynamical_2019,remund_semi-classical_2022,dahlbom_sunnyjl_2025,kim_emulation_2025} which however likely miss genuine quantum effects at small $S$.

Here we narrow this methodological gap by introducing a dynamic extension of the high-temperature expansion (HTE). Conventional HTE targets equal-time correlations and thermodynamic observables via graph-based expansions in $J/T$ and is developed for more than five decades \cite{dombPhaseTransitions1974,oitmaaSeriesExpansion2006}. It benefits from a formulation in the thermodynamic limit and is oblivious to frustration, entanglement, dimensionality and spin-length \cite{lohmannTenthorderHightemperature2014}. Its main limitation is the model-dependent temperature range $T \gtrsim O(J/4)$. Our dynamic HTE (Dyn-HTE) shares the same benefits but applies to frequency moments of the real-frequency dynamic susceptibility from which we reconstruct the DSF \cite{mori_continued-fraction_1965,lovesey_dynamic_1972,viswanath_recursion_1994}. Although these moments are considerably more complex than equal-time correlators, the high expansion orders of conventional HTE are maintained. This allows for the computation of four to seven frequency moments, depending on $T/J$, and goes beyond numerical linked-cluster expansions (NLCE) \cite{rigol_numerical_2007,tang_short_2013} which computed the lowest two moments (Gauss approximation) for some particular $S\!=\!1/2$ models \cite{sherman_structure_2018}.

Our open-source numerical implementation of Dyn-HTE \cite{Dyn-HTEsoftware_v1_0} makes it a versatile numerical tool catering directly to experiments. It currently covers $S\!\in\!\{1/2,1\}$ Heisenberg models with arbitrary geometry and a single $J$, for technical details see the companion paper \cite{burkard_DynHTE_long}. Here we derive the method and benchmark in 1D. We then obtain DSFs of the highly-frustrated triangular and pyrochlore lattice for $S\!=\!1/2$ and $S\!=\!1$, respectively. The former sheds new light on the nature of the model's anomalous intermediate-$T$ regime while the latter is in fair agreement with existing neutron scattering data. 

\emph{Model and DSF}.—We consider length-$S$ quantum spins $S^\alpha_i$ with $\alpha = x,y,z$ and ladder operators $S_i^{\pm}\!=\!\left( S_i^{x}\pm iS_i^{y} \right) /\sqrt{2}$ at sites $\mathbf{r}_i$ ($i=1,2,...,N$) of an arbitrary lattice $\mathcal{L}$. We focus on the Heisenberg model 
\begin{equation}
    H=J\sum_{(ii^\prime)}\left(S_{i}^{+}S_{i^\prime}^{-}+S_{i}^{-}S_{i^\prime}^{+}+S_{i}^{z}S_{i^\prime}^{z}\right), \label{eq:H}
\end{equation}
characterized by a \emph{single} coupling $J$ on arbitrary bonds $(ii^\prime)$ with $i<i^\prime$. This includes the important case of symmetry-related nearest-neighbor interactions on which we focus in the following. None of these assumptions or the spin-rotation symmetries of~\eqref{eq:H} are essential for Dyn-HTE and can be relaxed in future developments analogous to conventional HTE \cite{rosner_high-temperature_2003,oitmaaSeriesExpansion2006,lohmannTenthorderHightemperature2014,pierre_high_2024}. 
The DSF for translation invariant $\mathcal{L}$ is defined as
\begin{equation}
S(\mathbf{k},\omega) = \int_{-\infty}^{+\infty} \! 
\frac{\mathrm{d}t}{2\pi N}
        \sum_{i,i^\prime}  
    e^{i\omega t-i\mathbf{k}\cdot(\mathbf{r}_i-\mathbf{r}_{i^\prime})}
    \!
    \left\langle S_{i}^{z}(t)S_{i^\prime}^{z}\right\rangle\! \label{eq:DSF_lett}
\end{equation}
and relates to the imaginary part of the dynamical susceptibility via the fluctuation-dissipation relation \cite{bruusManyBodyQuantum2004}, $A_\mathbf{k}(\omega)=2\pi (1-e^{-\omega/T}) S(\mathbf{k},\omega)$. From the Lehmann representation, the exact $A_{\mathbf{k}}(\omega)$ fulfills $A_{\mathbf{k}}(\omega>0)\geq0$ and $A_{\mathbf{k}}(\omega)=-A_{-\mathbf{k}}(-\omega)$. We now assume inversion symmetry, then $A_{\mathbf{k}}(\omega)=A_{-\mathbf{k}}(\omega)$ is antisymmetric in $\omega$.

\emph{Continued fraction and moment expansion}.—It is well known \cite{mori_continued-fraction_1965,lovesey_dynamic_1972,viswanath_recursion_1994,pairault_strong-coupling_2000,perepelitsky_transport_2016} that such a \emph{faithful} $A_{\mathbf{k}}(\omega)$ can be generated from a continued fraction expansion, 
\begin{eqnarray}
    TA_{\mathbf{k}}(\omega) &=& 2 \,w \,\mathrm{Re}[\tilde{R}_{\mathbf{k}}(s=iw+0^+)],    \label{eq:S_and_A_from_R} \\ 
    \tilde{R}_\mathbf{k}(s) &=& \frac{\delta_{\mathbf{k},0}}{s+}\,\frac{\delta_{\mathbf{k},1}}{s+}\,\frac{\delta_{\mathbf{k},2}}{s+}\,\frac{\delta_{\mathbf{k},3}}{s+}\cdots\, , \label{eq:contFrac}
\end{eqnarray}
with $w=\omega/J$ the dimensionless frequency and the continued fraction parameters $\delta_{\mathbf{k},r} \geq 0$. The first $\delta_{\mathbf{k},r}$ for $r \leq r_{max}$ can be determined \cite{mori_continued-fraction_1965,viswanath_recursion_1994} from the first $r_{max}$ (even) frequency-moments of the relaxation function $R_{\mathbf{k}}(w) \! \equiv \! TA_\mathbf{k}(\omega)/(2\pi w)$ \cite{lovesey_dynamic_1972},
\begin{equation} \label{eq:moments}
    m_{\mathbf{k},2r}=\int_{-\infty}^{\infty} \! \! \mathrm{d}w \,\, w^{2r}R_{\mathbf{k}}(w), \;\; r=0,1,2,...\, .
\end{equation} 
Dropping the $\mathbf{k}$-subscript for brevity, the connection is made via Laplace transform of Eq.~\eqref{eq:moments},
$\tilde{R}(s)=\int_{-\infty}^{\infty}\mathrm{d}v\,\frac{R(v)}{s-iv} $
and expansion in the inverse complex argument, $1/s$. Comparison of coefficients \cite{viswanath_recursion_1994} leads to
$\delta_{0}\!=\!m_{0}$, $\delta_{1}\!=\!m_{2}/m_{0}$, $\delta_{2}  =  m_{4}/m_{2}-m_{2}/m_{0}$ and
$
\delta_{3}\!=\!m_{0}\left(m_{4}^{2}-m_{2}m_{6}\right)/[m_{2}^{3}-m_{0}m_{2}m_{4}]$.
For $\delta_{4,5,6}$ required in the following, see End Matter.
Via a short-time expansion of the spin correlator, the moments can be expressed as equal-time correlators involving a $2r-1$-fold commutator, $J^{2r}m_{\mathbf{k},2r}/(2T)=M_{\mathbf{k},2r-1}$ defined as the spatial Fourier transform of
\begin{equation}
M_{ii^{\prime},n}
\!\!=\! 
i^{n}\partial_{t}^{n}
\!\!
\left\langle S_{i}^{z}(t)S_{i^{\prime}}^{z}\right\rangle \! |_{t=0}
\!=\!\!
\left\langle [...[[S_{i}^{z},\!H],\!H],\!...,\!H]S_{i^{\prime}}^{z}\right\rangle  \!.\!\! \label{eq:momentFromCommutators} 
\end{equation}
As exact calculation of all $m_{\mathbf{k},2r}$ is not possible for general models one is limited to finite $r \leq r_{max}$. We discuss an extrapolation strategy on the level of the $\delta_{\mathbf{k},r}$ below.

Inspection of Eq.~\eqref{eq:momentFromCommutators} shows that the amount of computable moments $r_{max}$ strongly depends on dimension (controlling the growth of operator support under application of $[...,H]$), spin length $S$ and temperature. The most simple case is $T=\infty$ where equal-time spin correlators factorize according to site indices. Here, depending on the concrete model, $r_{max}=O(30)$ can be reached in 1D \cite{jung_transport_2006,wangDiffusionConstantsRecursion2024} while $r_{max} \lesssim 10 $ in 3D \cite{viswanath_recursion_1994}. However, the case of finite $T$ is complicated by the presence of the thermal density matrix $\rho= \! e^{-H/T}\!/\mathrm{tr}[e^{-H/T}]$ in Eq.~\eqref{eq:momentFromCommutators}. To the best of our knowledge, no method to evaluate even a moderate number of moments at $T/J=O(1)$ for possibly frustrated Heisenberg models beyond 1D is established.

\emph{Dyn-HTE for moments}.—We now present our main conceptual result relevant for finite $T$: The HTE of moments $m_{\mathbf{k},2r}$ is encoded in the Dyn-HTE of the Matsubara spin correlator. This is significant as the latter treats $\rho$ and imaginary-time evolution $S_{i}^{z}(\tau) = e^{H\tau}S_{i}^{z}e^{-H\tau}$ on equal footing \cite{bruusManyBodyQuantum2004} and allows for an efficient evaluation of \emph{exact} HTE coefficients, see our companion article \cite{burkard_DynHTE_long}. The Matsubara correlator at frequency $\nu_m = 2 \pi T m$ \!($m  \in  \mathbb{Z}$) and its HTE to order $n_{max}$ in $x \!=\!  J/T$ read
\begin{eqnarray}
    && \!TG_{ii^\prime}(i\nu_{m}) \! =T^2 \!\!\! \int_{0}^{1/T} \!\!\!\!\!\! \mathrm{d}\tau\mathrm{d}\tau^{\prime}\,e^{i\nu_{m}(\tau-\tau^{\prime})} \!
\left\langle \mathcal{T}S_{i}^{z}(\tau)S_{i^{\prime}}^{z}(\tau^{\prime})\right\rangle
\label{eq:GMatsubara} \\ 
&&=p_{ii^{\prime}}^{(0)}(x)\delta_{0,m}\!+\!\!\sum_{r=1}^{r_{max}}p_{ii^{\prime}}^{(2r)}(x) x^{2r} \!\Delta_{2\pi m}^{2r} \! + \! O(x^{n_{max}+1}).  \label{eq:GMatsubara_HTE}
\end{eqnarray}
Here $\mathcal{T}$ enforces $\tau$-ordering, $\Delta_{2\pi m} \equiv (1-\delta_{0,m})/(2\pi m)$ and $r_{max}=\left\lfloor n_{max}/2\right\rfloor$. The polynomials $p_{ii^{\prime}}^{(2r)}(x)$ will be linked to the HTE of the moments momentarily. They are of degree $n_{max}-2r$ with rational coefficients. Our open-source implementation \cite{Dyn-HTEsoftware_v1_0} for $n_{max}=12$ provides $p_{ii^{\prime}}^{(2r)}(x)$ for all site-pairs $ii^\prime$ on \emph{arbitrary} lattices $\mathcal{L}$.
This flexibility hinges on the pre-evaluation of Dyn-HTE \eqref{eq:GMatsubara_HTE} on generic lattice \emph{snippets} called ``graphs" which are then embedded in arbitrary (and also high-dimensional and frustrated) $\mathcal{L}$ in a post-processing step as in conventional HTE \cite{oitmaaSeriesExpansion2006}. Efficient evaluation of \emph{all} $\sim \! 10^6$ graphs with up to $n_{max}$ edges [currently for $S\in \{\frac{1}{2},1\}$] rests on a
recursive formulation of perturbation theory and the exact solution of the $n_{max}\!+\!2$-fold $\tau$-integrals using the kernel trick~\cite{halbingerSpectralRepresentation2023}, see Ref.~\onlinecite{burkard_DynHTE_long} for details.
There we also discuss results for the \emph{static} spin susceptibility, $\sim G_{ii^\prime}(i\nu_m=0)$.

We show that the spatial Fourier transform of the polynomials $p_{ii^{\prime}}^{(2r)}(x)$ in Eq.~\eqref{eq:GMatsubara_HTE} are identical to the HTE of the relaxation function's frequency moments $m_{\mathbf{k},2r}$ in Eq.~\eqref{eq:moments} [for $r>0$, there is a relative sign $-(-1)^r$].  To see this, consider the standard integral-relation between $A_{\mathbf{k}}(\omega)$ and the Matsubara correlator in momentum space \eqref{eq:GMatsubara},
\begin{equation}
G_{\mathbf{k}}(i\nu_{m}) = \frac{1}{2\pi}\int_{-\infty}^{\infty}\mathrm{d}\omega \, \frac{A_{\mathbf{k}}(\omega)}{\omega-i\nu_{m}}. \label{eq:GMatsubaraVsA}
\end{equation}
This is strictly correct only for $m \neq 0$ and an additional term quantifying long-term memory effects and non-ergodicity appears on the right-hand side for $m = 0$ \cite{kwok_correlation_1969,suzuki_ergodicity_1971,watzenbock_long-term_2022}. For systems of interest in this work we assume such effects are absent and thus continue with Eq.~\eqref{eq:GMatsubaraVsA}. The claim follows from an expansion of the right-hand side of Eq.~\eqref{eq:GMatsubaraVsA} in (even) powers of $1/\nu_m \sim \Delta_{2\pi m}$,
\begin{equation}
    TG_{\mathbf{k}}(i\nu_{m})\!=\!\!\int_{-\infty}^{\infty}
    \!\!\!\!
    \mathrm{d}w \,R_{\mathbf{k}}(w)
    \Bigl[\delta_{m,0}
    -\!\!\!\!\!
    \sum_{r=1,2,...}
    \!\!\!
    (iwx\Delta_{2\pi m})^{2r}
    \Bigr],\label{eq:analytic-cont_expanded}
\end{equation}
after comparison to Eq.~\eqref{eq:moments} and \eqref{eq:GMatsubara_HTE}.

To lift the assumption of inversion symmetry (under which $A_{\mathbf{k}}(\omega)$ is antisymmetric in $\omega$ and $G_{\mathbf{k}}(i\nu_{m})$ is real \cite{burkard_DynHTE_long}), the above can be straightforwardly generalized using the symmetric and anti-symmetric combination $A_{\mathbf{k}}^{\pm}(\omega)\equiv\frac{1}{2}\left[A_{\mathbf{k}}(\omega)\mp A_{-\mathbf{k}}(-\omega)\right]$
which fulfill the analog of Eq.~\eqref{eq:GMatsubaraVsA} with the imaginary and real part of $G_{\mathbf{k}}(i\nu_{m})$ on the left-hand side, respectively.

\begin{figure}
\begin{centering}
\includegraphics{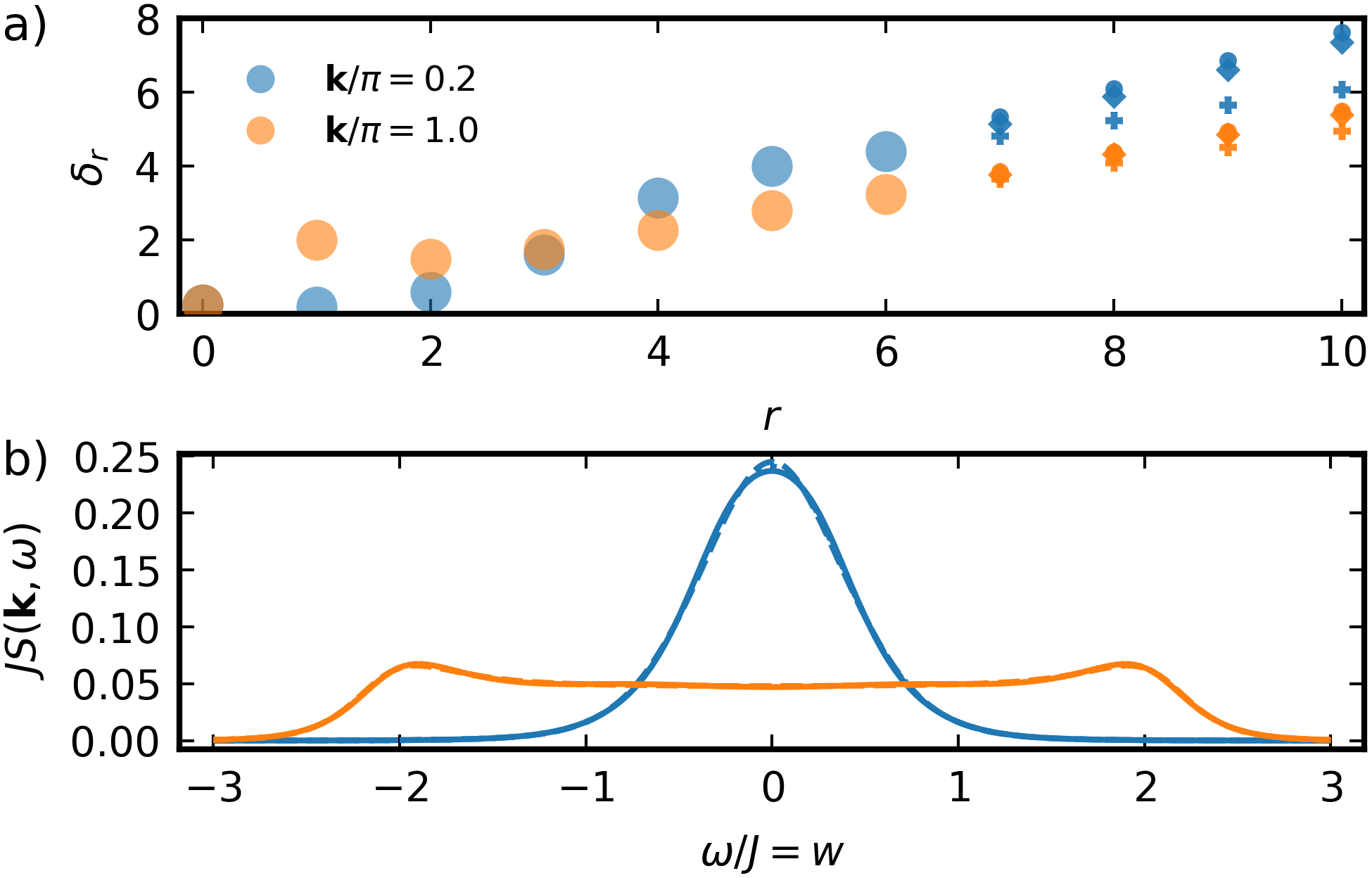}
\par\end{centering}
\centering{}\caption{\label{fig:Chain_delta_r_Tinf} 
Heisenberg $S=1/2$ AFM chain at $T=\infty$. (a) Continued fraction parameters $\delta_r$ for two momenta $\mathbf{k}$. The large dots denote the exact results for $r=0,1,...,6$ from Dyn-HTE, the small markers depict various linear extrapolation schemes $\delta_{r>6} = (r-6)a + b$ to which the DSF in (b), obtained from the infinite continued fraction Eq.~\eqref{eq:linear_Approx}, is largely insensitive as seen from the overlap of various linestyles.
}
\end{figure}
\begin{figure*}
\begin{centering}
\includegraphics[width=1.0\textwidth]{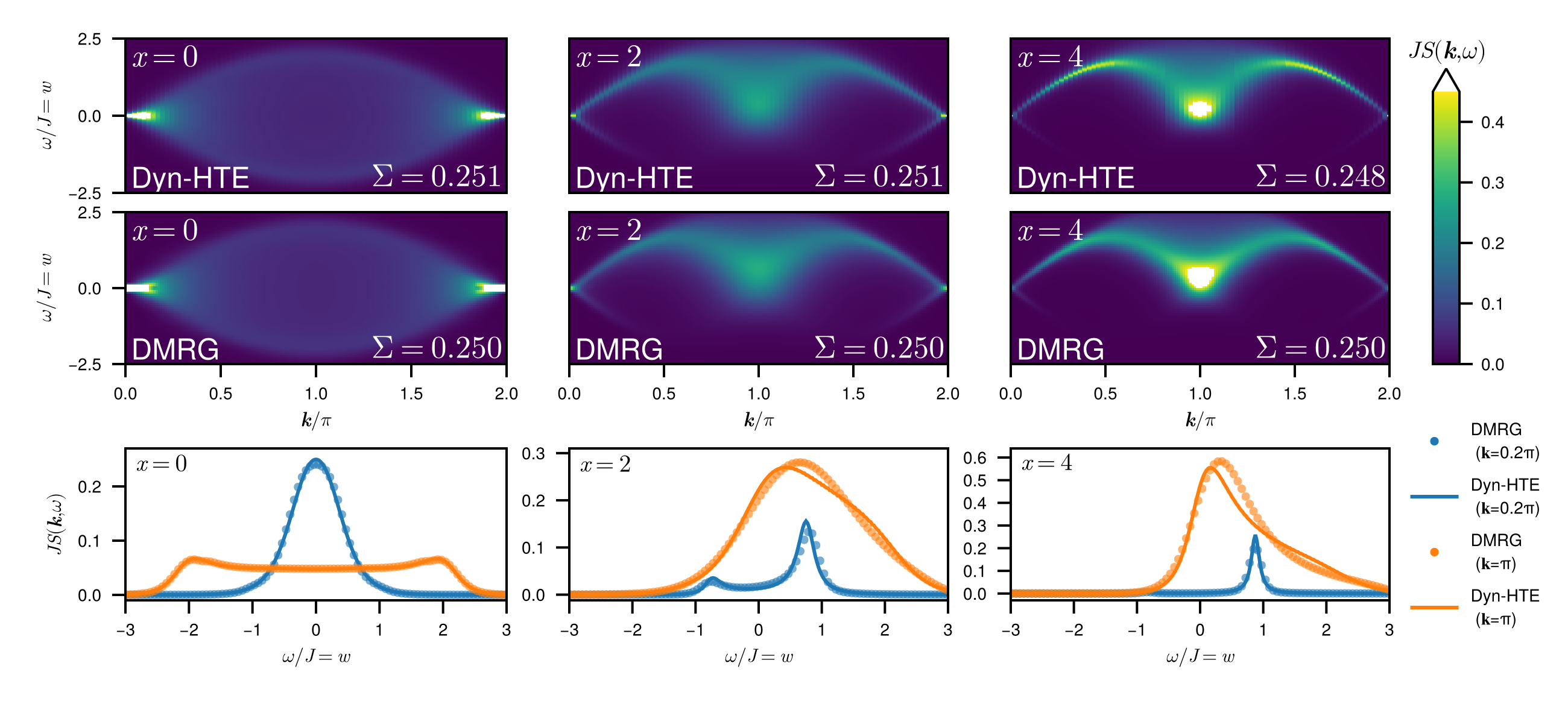}
\par\end{centering}
\centering{}\caption{\label{fig:chain_dynHTE_DMRG_comparison} 
DSF for the Heisenberg $S=1/2$ AFM chain at $x=J/T \in \{0,2,4\}$, left to right column. Top row: Dyn-HTE results based on extrapolation of moments with $r\leq6$ ($r\leq3$ for $x>0$). Middle row: DMRG data reproduced from Ref.~\onlinecite{kish_high-temperature_2024}. Bottom row: Lineshape for $S(\mathbf{k},\omega)$ at fixed $\mathbf{k}\in \{0.2 \pi, \pi\}$. The lattice spacing is set to unity. The values for
$
    \Sigma \equiv \frac{J}{V_{BZ}}  
    \! \int_{-\infty}^\infty \! \mathrm{d} \omega \!
    \int_{BZ} \! \mathrm{d}\mathbf{k} \:S(\mathbf{k},\omega)
$
with $V_{BZ}$ the BZ-volume are also indicated and Dyn-HTE fulfills the sum-rule $\Sigma=\langle S_i^z S_i^z \rangle  = S(S+1)/3=1/4$ within $<1\%$.}
\end{figure*}
\emph{Heisenberg $S=1/2$ AFM chain}.—As a first example, we turn to the well-studied nearest-neighbor Heisenberg $S=1/2$ anti-ferromagnet (AFM) in 1D for benchmark.
Due to finite expansion order $n_{max}=12$, the HTE is only available for moments $m_{\mathbf{k},2r}$ with $r \leq r_{max}=6$. To show how to deal with this restriction, we first consider the infinite-$T$ case ($x=0$) where moments are \emph{exactly} given by the $x^0$-coefficients of their HTEs. We convert to the $\delta_r$ and plot them in Fig.~\ref{fig:Chain_delta_r_Tinf}(a) for two representative momenta $\mathbf{k}=0.2\pi$ and $\pi$ (large dots). To avoid sharp Dirac-Delta peaks in $A_\mathbf{k}(\omega)$ resulting from an abrupt termination of the continued fraction \eqref{eq:contFrac}  we follow Ref.~\onlinecite{viswanath_recursion_1994} and linearly extrapolate $\delta_{r>r_{\max}} = (r-r_{\max}) a +b$  for $ r>r_{max} $. Then Eq.~\eqref{eq:contFrac} can be written as 
\begin{equation}
 \tilde{R}_{r_{max}}(s)=\frac{\delta_{0}}{s+}\,\frac{\delta_{1}}{s+}\,\frac{\delta_{2}}{s+}\,\cdots \,\frac{\delta_{r_{max}-1}}{s+\delta_{r_{max}} \Gamma_{a,b}(s)}, \label{eq:linear_Approx}
\end{equation}
where the termination function $\Gamma_{a,b}(s)$ is a fraction of Hermite polynomials \cite{cuyt2008handbook} (see End Matter) and allows to take the limit $0^+$ in Eq.~\eqref{eq:S_and_A_from_R}  exactly.
The different small markers in Fig.~\ref{fig:Chain_delta_r_Tinf}(a) correspond to various linear extrapolation schemes (c.f.~figure caption). The main features of the resulting DSFs shown in various linestyles in Fig.~\ref{fig:Chain_delta_r_Tinf}(b) are practically insensitive to the chosen extrapolation and is in excellent agreement with DMRG~\cite{kish_high-temperature_2024} throughout the Brillouin zone (BZ), see Fig.~\ref{fig:chain_dynHTE_DMRG_comparison}, left column.

We now turn to $T<\infty$ $(x>0)$ where Dyn-HTE has its main advantage. Recall that it provides the expansion of moments $m_{\mathbf{k},2r}(x)$ in $x$ to order $n_{max}-2r$. Hence, for the interesting regime $x \gtrsim 1$ beyond convergence of the bare HTE series, we apply resummation in form of Padé approximants \cite{oitmaaSeriesExpansion2006} to $m_{\mathbf{k},2r}(x)$ with $r\leq 3$ and then proceed as above, see End Matter (Fig.~\ref{fig:chain_moments_and_deltas_finiteT}) for details. Analogous to conventional HTE \cite{elstner_spin-12_1994,oitmaaSeriesExpansion2006} we improve convergence by changing to the auxiliary variable $u=\tanh(f x)$ with tuning parameter $f$ chosen such that different Padé approximants agree. In principle, $f$ should be as small as possible to reach large $x$ and could vary with $\mathbf{k}$ but for the chain a single $f=0.48$ works well. The 2nd and 3rd column of Fig.~\ref{fig:chain_dynHTE_DMRG_comparison} compare the DSF at temperatures $x=2$ and $4$ from Dyn-HTE (top) to DMRG \cite{kish_high-temperature_2024} (middle). Linecuts (bottom) show only minor deviations.

\emph{$S=1/2$ Heisenberg AFMs in 2D}.—While magnetic long-range order is only possible at $T=0$ for nearest-neighbor models, its precursor in the form of spin-waves can dominate the DSF already at $T \lesssim J$ \cite{chakravartyTwodimensionalQuantum1989}. We refer to the End Matter for representative Dyn-HTE results for the DSF of the square lattice model and the frustrated kagome lattice which lacks long-range magnetic order at $T=0$ and shows broad features in the DSF.

Here we focus on the triangular lattice model with $120^\circ$-order at $T=0$ \cite{capriotti_long-range_1999}. For $ \frac{1}{4} \lesssim T/J \lesssim 1$, however, an enigmatic anomaly occurs: As first found by HTE, static properties deviate strongly from renormalized-classical spin-wave predictions \cite{elstner_finite_1993, ghioldi_dynamical_2018,chen_two-temperature_2019}. For example, at $T=J/4$, the correlation length is relatively small (about a lattice constant) and the entropy per spin is large ($\simeq\frac{\mathrm{ln}2}{3}$).

\begin{figure}
\includegraphics{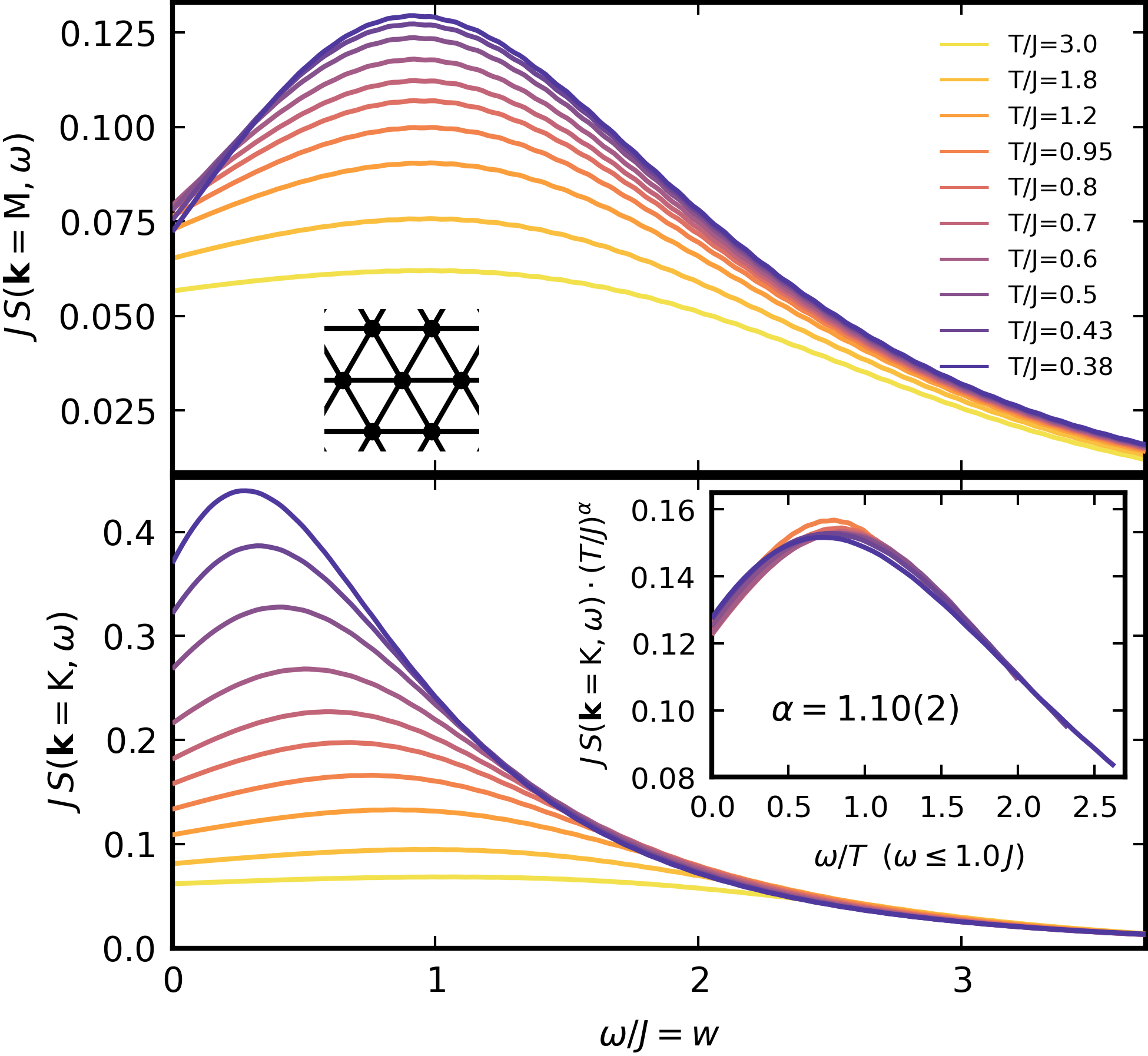}
\caption{\label{fig:Triangular} 
DSF of triangular lattice $S=1/2$ Heisenberg AFM at momenta $M$ (top) and $K$ (bottom) from Dyn-HTE with $r_{max}=3$ and $f=0.55$. Inset: $\omega/T$-scaling collapse for $\omega \leq J$ and $0.43 \leq T/J \leq 0.95$ with scaling exponent $\alpha=1.10(2)$.
}
\end{figure}

The DSF from Dyn-HTE shown in Fig.~\ref{fig:Triangular} (see also End Matter for a BZ-path at $x=3$) is ideally suited to shed new light on the still debated nature of this intermediate-$T$ regime. One proposed scenario starts from the $T=0$ excitation spectrum which shows a ``roton-like excitation" (RLE) characterized by a dispersion minimum at the $M$-point (center of BZ edge) with gap $\Delta \simeq 0.55J$ \cite{zheng_anomalous_2006,starykh_flat_2006}. Is the intermediate-$T$ region primarily characterized by thermal excitations of these enigmatic RLEs? At $\mathbf{k}=M$, the $T$-dependence of the equal-time structure factor shows a weak maximum around $T\simeq \Delta$, see Ref.~\onlinecite{chen_two-temperature_2019}, but the DSF in Fig.~\ref{fig:Triangular}(top) does not soften significantly across the intermediate $T$-range with the peak remaining around $\omega_{max}\simeq J$ [for $S(\mathbf{k}\!=\!M,\omega \!\rightarrow \!0)$, a weak maximum can however be found as $T$ decreases]. Hence, if the RLE at $\mathbf{k}=M$ is responsible for the anomalous behavior, the associated fluctuations are not simple spin-waves but must be more complex. One candidate are triangle-based chiral fluctuations \cite{chen_two-temperature_2019} but the associated chiral DSF is beyond scope of this work. 

An alternative scenario links the anomalous intermediate-$T$ regime to passage through a critical fan \cite{sachdev_quantum_2011} of a $T=0$ quantum phase transition (QPT) at which the $120^\circ$ order melts \cite{sachdev_universal_1992,chubukov_quantum_1994,ghioldi_dynamical_2018}.  A hallmark of this scenario is a temperature-frequency scaling relation for the DSF at ordering wavevector $\mathbf{k}\!=\!K$ for $|\omega| \lesssim J$,
\begin{equation}
JS(\mathbf{k}=K,\omega)(T/J)^\alpha=\Phi(\omega/T), 
    \label{eq:triangular_scaling}
\end{equation}
where $\alpha=(2-\eta)/z$ is determined by the critical exponents of the QPT and $\Phi(\cdot)$ is a scaling function. Interestingly, the $\mathbf{k}=K$ DSF from Dyn-HTE shown in Fig.~\ref{fig:Triangular}(bottom) fulfills such a relation for $\alpha=1.10(2)$ (inset), but the $T$-range for which scaling occurs is relatively narrow. One candidate for the QPT in question is the transition to a Dirac QSL in the $J_1$-$J_2$-model at $J_2 \simeq 0.06 J_1$ \cite{zhu_spin_2015}. Although the critical exponents are unknown, Lorentz invariance would imply $z=1$ and $\eta = O(1)$ has been found in the potentially related Gross-Neveu Heisenberg model via large-$N_f$ or $\epsilon=4-d$ expansion \cite{zerfCriticalPropertiesNeel2019,dupuisTransitionDiracSpin2019}. Our value of $\alpha$ is thus not implausible and can inform further efforts on the field-theory side, although the scaling analysis should be repeated with Dyn-HTE data for the $J_1$-$J_2$-model.
Other open questions pertain to the incompatible value of $\alpha\simeq 1.73(12)$ found from a similar scaling analysis of INS data from $\mathrm{KYbSe}_2$ with finite and slightly subcritical $J_2$ \cite{scheie_proximate_2024} and an extended scaling analysis taking into account deviations from the ordering wavevector, $\mathbf{k}\neq K$.

\begin{figure}
    \vspace{-0.7cm}
    \centering
    \includegraphics[width= 1.0\linewidth]{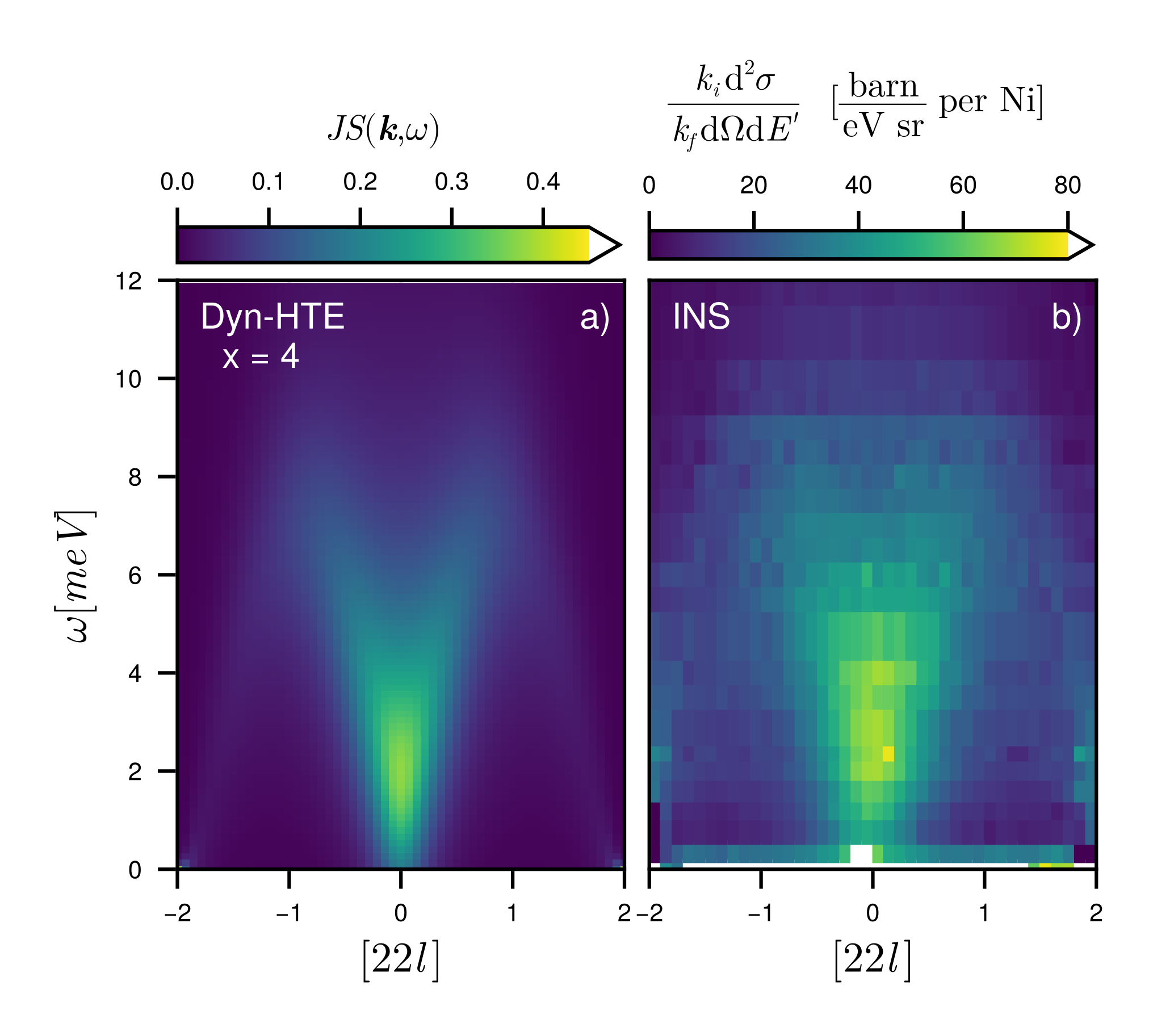}
     \vspace{-1.2cm}
    \caption{(a) DSF for the pyrochlore lattice $S=1$ Heisenberg AFM for $\mathbf{k} = (2,2,l)$, $J=\SI{2.4}{\meV}$ at $x=J/T=4$ from Dyn-HTE (via u-Padé with $f=0.6$). (b) Experimental INS data for $\mathrm{NaCaNi}_2\mathrm{F}_7$ at $x\simeq 15$, from Ref.~\onlinecite{plumb_continuum_2019}.}
    \label{fig:PyrochloreINS}
\end{figure}

\emph{$S\!=\!1$ Heisenberg AFMs: Chain and pyrochlore lattice}.—Finally, we obtain the DSF from Dyn-HTE for spin $S=1$ Heisenberg AFMs. For the chain at $x=4$, some aspects like the dispersion maxima or Haldane gap already agree well with ground-state ($x \! \rightarrow \! \infty$) DMRG results, see End Matter (Fig.~\ref{fig:S1_chain_DSF}). 
The $S=1$ pyrochlore material $\mathrm{NaCaNi}_2\mathrm{F}_7$ 
was reported to approximately realize the nearest-neighbor Heisenberg AFM \cite{zhang_dynamical_2019,plumb_continuum_2019} with $J=\SI{2.4}{\meV}$ \cite{pohle2025abundance}. In Fig.~\ref{fig:PyrochloreINS}, we compare the DSF from Dyn-HTE along the $[22l]$ momentum with experimental INS results \cite{plumb_continuum_2019}. We find fair agreement of the main features including the size of the ``V" shape as well as the height of the dome-shape which was not captured in semiclassical studies \cite{pohle2025abundance,zhang_dynamical_2019}. In Dyn-HTE, the maximum along the line-cut at $[220]$ appears at slightly lower $\omega$ compared to INS data. Possible reasons are potential beyond-Heisenberg perturbations in the material and a mismatch in temperature ($x=4$ in Dyn-HTE versus~$x\simeq 15$ in INS). Indeed the peak position in Dyn-HTE has a trend to higher $\omega$ with lower $T$.

\emph{Conclusion}.—Compared to static observables of conventional HTE, the DSF obtained via Dyn-HTE accesses a wealth of additional information on dipolar excitations down to moderately-low $T \gtrsim J/4$. On a technical level we achieve this by exploiting a graph-based approach for high-order moment expansions and fine momentum resolution, combined with continued-fraction reconstruction, judiciously chosen resummation schemes all packaged in a single algorithmic recipe \cite{Dyn-HTEsoftware_v1_0}. This combination sets our method apart from previous (spin-)diagrammatic approaches in a qualitative manner \cite{vaksSelfconsistentField1967,starykhDynamicsSpinHeisenberg1997,ruckriegel_recursive_2024,SchneiderDipolarOrdering2024,khatamiLinkedclusterExpansionGreens2014,perepelitsky_transport_2016}.

Looking ahead, Dyn-HTE will not only be useful to model INS data but also informs experiments with local probes like muon spin relaxation \cite{khatua_experimental_2023}, inelastic electron tunneling- or noise-spectroscopy for 2D quantum magnets \cite{feldmeier_local_2020,konig_tunneling_2020,davis_probing_2023}. Other applications include finite-$T$ spin diffusion and the computation of quantum Fisher information as entanglement witness \cite{hauke_measuring_2016,kish_high-temperature_2024}.
Straightforward extensions of the Dyn-HTE formalism will consider advanced resummation strategies \cite{kim_homotopic_2021}, $S>1$, models with multiple couplings (e.g.~$J_1$-$J_2$) \cite{hehnHightemperatureSeries2017}, single-ion anisotropies or magnetic fields \cite{pierre_high_2024}. Another worthwhile extension would be higher-order dynamic \cite{kaib_nonlinear_2025,watanabe_exploring_2024} or chiral \cite{chen_two-temperature_2019,bornet_enhancing_2024,ruckriegel_recursive_2024} correlators moving into focus recently.

\begin{acknowledgments}
\emph{Acknowledgments}.—We thank 
Michel Ferrero, Matías Gonzalez, Lukas Janssen, Wei Li, Johannes Reuther, Alexander Tsirlin, Andreas Weichselbaum, and Kemp Plumb for useful discussions and the latter two for sharing DMRG data from Ref.~\onlinecite{kish_high-temperature_2024} and INS data from Ref.~\onlinecite{plumb_continuum_2019}, respectively.
The authors acknowledge the Gauss Centre for Supercomputing e.V. (www.gauss-centre.eu) for funding this project by providing computing time through the John von Neumann Institute for Computing (NIC) on the GCS Supercomputer JUWELS at Jülich Supercomputing Centre (JSC). The authors also acknowledge support by the state of Baden-Württemberg through bwHPC and the German Research Foundation (DFG) through Grant No.~INST 40/575-1 FUGG (JUSTUS 2 cluster).
We acknowledge funding from the Deutsche
Forschungsgemeinschaft (DFG, German Research Foundation) through the Research Unit FOR 5413/1, Grant No.~465199066. B.Sch.~acknowledges funding from the Munich
Quantum Valley, supported by the Bavarian state government
with funds from the Hightech Agenda Bayern Plus. B.Sb. and
B.Sch.~are supported by DFG Grant No.~524270816.
\end{acknowledgments}

\bibliography{DynHTE.bib}

\clearpage
\appendix
\onecolumngrid
\begin{center}
    \textbf{\large End Matter} 
\end{center}

\emph{Continued-fraction parameters:} The parameters $\delta_r$ for $r=4,5,6$ are given in terms of moments $m_0,m_2,...,m_{2r}$:
\begin{eqnarray}
\delta_4 &=& \frac{m_2 \left(m_4^3-(2 m_2 m_6+m_0 m_8) m_4+m_0 m_6^2+m_2^2 m_8\right)}{\left(m_2^2-m_0 m_4\right) \left(m_2 m_6-m_4^2\right)} \\
\delta_5 &=& -\frac{\left(m_2^2-m_0 m_4\right) \left(-m_6^3+2 m_4 m_8 m_6-m_2 m_8^2+\left(m_2 m_6-m_4^2\right) m_{10}\right)}{\left(m_4^2-m_2 m_6\right) \left(m_4^3-(2 m_2 m_6+m_0 m_8) m_4+m_0 m_6^2+m_2^2 m_8\right)} \\
\delta_6 &=& [(m_4^2-m_2 m_6)((3 m_4 m_8+2 m_2 m_{10}+m_0 m_{12}) m_6^2-m_6^4-2 \left(\left(m_4^2+m_0 m_8\right) m_{10}+m_2 \left(m_8^2+m_4 m_{12}\right)\right) m_6\\
&+&m_0 m_8^3-m_4^2 m_8^2-m_2^2 m_{10}^2+m_0 m_4 m_{10}^2+2 m_2 m_4 m_8 m_{10}+\left(m_4^3+\left(m_2^2-m_0 m_4\right) m_8\right) m_{12})]\nonumber\\
&/&[\left(m_4^3-(2 m_2 m_6+m_0 m_8) m_4+m_0 m_6^2+m_2^2 m_8\right) \left(m_6^3-2 m_4 m_8 m_6+m_2 m_8^2+\left(m_4^2-m_2 m_6\right) m_{10}\right)]\nonumber
\end{eqnarray}

\emph{Termination function:} The infinite continued fraction with linearly growing parameters $
\Gamma_{a,b}(s)=\frac{1}{s+}\,\frac{(1 a+ b)}{s+}\,\frac{(2 a +b)}{s+}\,\frac{(3a +b)}{s+}\cdots \, . \label{eq:contFrac_lin} $
with $a,b>0$ fulfills the functional equation $1/\Gamma_{a,b}(s) = s + ( 1 a + b)\Gamma_{a,1a +b}(s)$ which is solved by a fraction of Hermite polynomials $H_{\nu}(z)$: $
\Gamma_{a,b}(s)=\sqrt{2/a}H_{-1-b/a}\left(s/\sqrt{2a}\right)/H_{-b/a}\left(s/\sqrt{2a}\right)
$. The special case $b\!=\!0$ yields $ \Gamma_{a,0}(s) \!=\! \sqrt{\frac{\pi }{2a}} e^{\frac{s^2}{2 a }} [1\!-\!\text{erf}(\frac{s}{\sqrt{2a
   }})]$ and analytically continues to a Gaussian  $\mathrm{Re}[\Gamma_{a,0}(i w+ 0^+)] = \sqrt{\frac{\pi }{2a}} e^{-\frac{w^2}{2 a }}$.

\begin{figure*}[b]
\begin{centering}
\vspace{-2mm}
\includegraphics[width=0.90\linewidth]{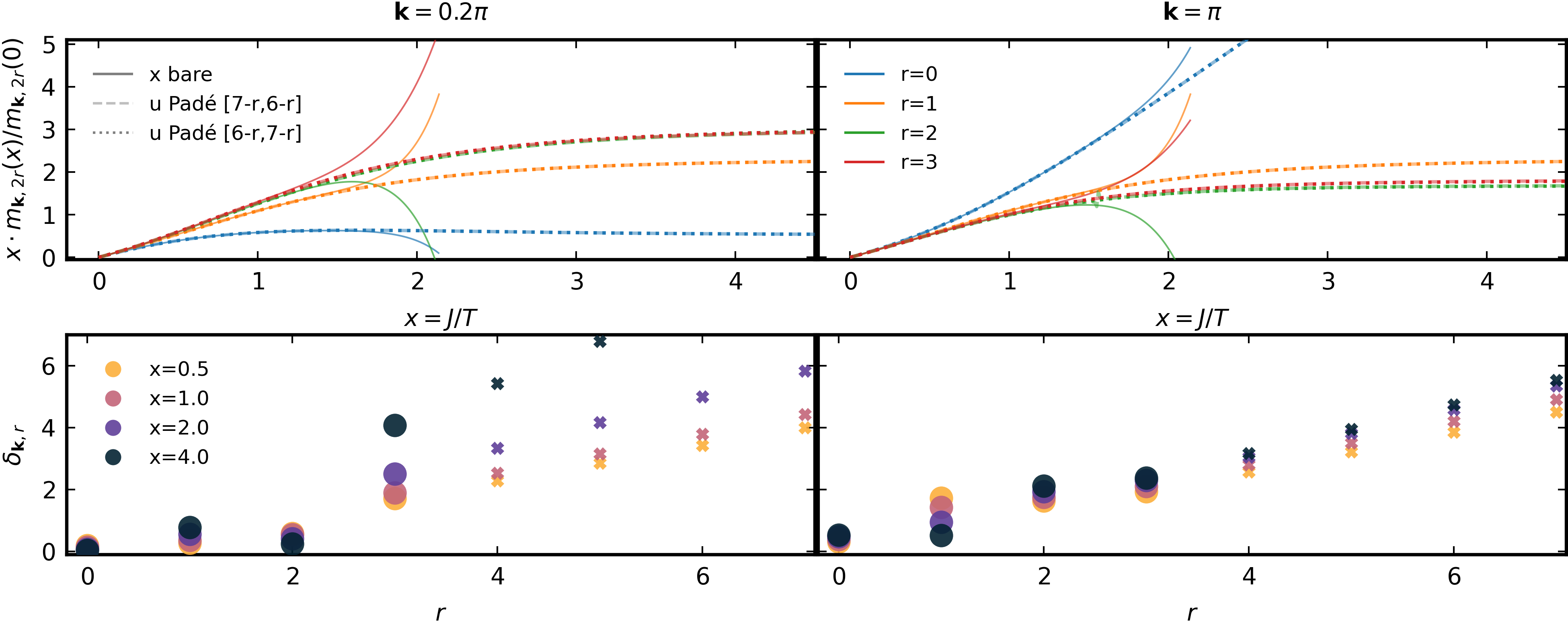}
\par\end{centering}
\centering{}\caption{\label{fig:chain_moments_and_deltas_finiteT} 
Heisenberg $S=1/2$ AFM chain: First four normalized moments $x \, m_{\mathbf{k},2r}(x)/m_{\mathbf{k},2r}(0)$, $r=0,1,2,3$ from Dyn-HTE (top) and corresponding continued fraction parameters $\delta_{\mathbf{k},r}$ (bottom) at momenta $\mathbf{k}=0.2 \pi$ (left) and $\mathbf{k}=\pi$ (right). The solid lines represent the bare series, while the dotted and dashed lines are different Padé approximants in the variable $u= \tanh(fx)$. We chose $f=0.48$ such that the different Padé approximants agree with each other. The bottom row shows the corresponding $\delta_r$ of the continued fraction expansion at various temperatures (dots) and their linear extrapolation for $r\geq 4$ (crosses).
}
\end{figure*}

\emph{Moments and continued fraction parameters for the Heisenberg $S=1/2$ AFM chain at $x>0$:} In Fig.~\ref{fig:chain_moments_and_deltas_finiteT} we provide details about the resummation of the Dyn-HTE moments $m_{\mathbf{k},2r}(x)$ for $x>0$ and the parameters $\delta_{\mathbf{k},r}$ underlying the DSF in Fig.~\ref{fig:chain_dynHTE_DMRG_comparison}. We consider $\mathbf{k}=0.2\pi$ and $\mathbf{k}=\pi$ in the left and right column, respectively. We find it convenient to show $x \, m_{\mathbf{k},2r} / m_{\mathbf{k},2r}(0)$ (top row), the factor $x$ cancels the $T$ which appears in the definition of the relaxation function and we thus expect a constant value for $x\rightarrow \infty$. The bare HTE series of $x \, m_{\mathbf{k},2r} / m_{\mathbf{k},2r}(0)$ is a polynomial of degree $1+n_{max}-2r=13-2r$ which diverges around $x=2$ (thin solid lines). Resummation uses Padé approximants after a change of variable $u= \tanh(fx)$, for the chain $f=0.48$ shows good agreement between different Padé approximants. Due to the saturation of $\tanh(\cdot)$, $x$ should not be taken larger than $O(2/f)$. In the bottom row we show the associated $\delta_{\mathbf{k},0},...,\delta_{\mathbf{k},3}$ and the linear extrapolation for $\delta_{\mathbf{k},4}$ and higher (via a line through $\delta_{\mathbf{k},3}$ and the origin). Note that all $\delta_{\mathbf{k},r}$ are non-negative as required for a faithful $A_\mathbf{k}(\omega)$, but this might break for poor moment resummations.

\begin{figure*}[ht!]
\begin{centering}
\hspace{-5mm}
\includegraphics[scale=0.094]{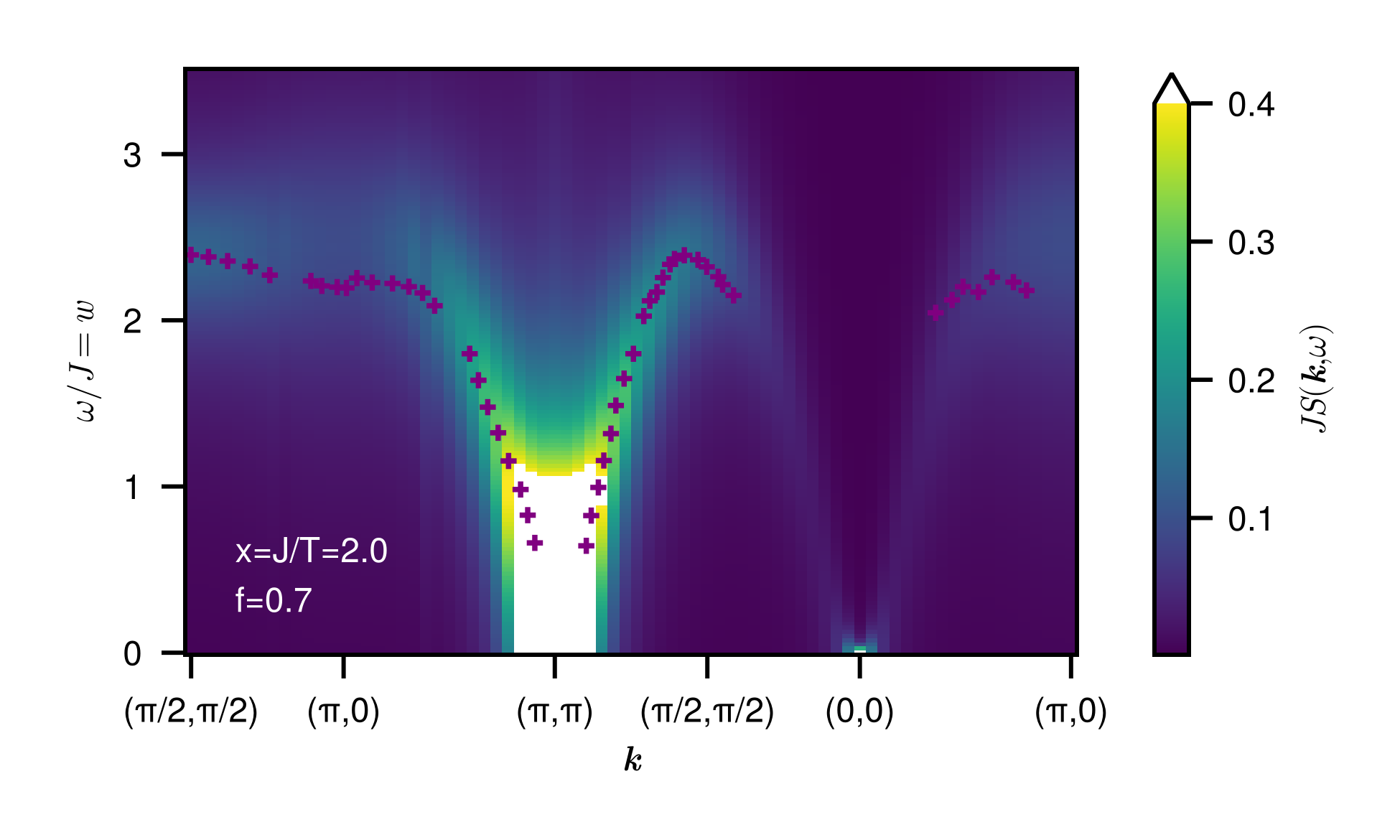}
\hspace{-3mm}
\includegraphics[scale=0.85]{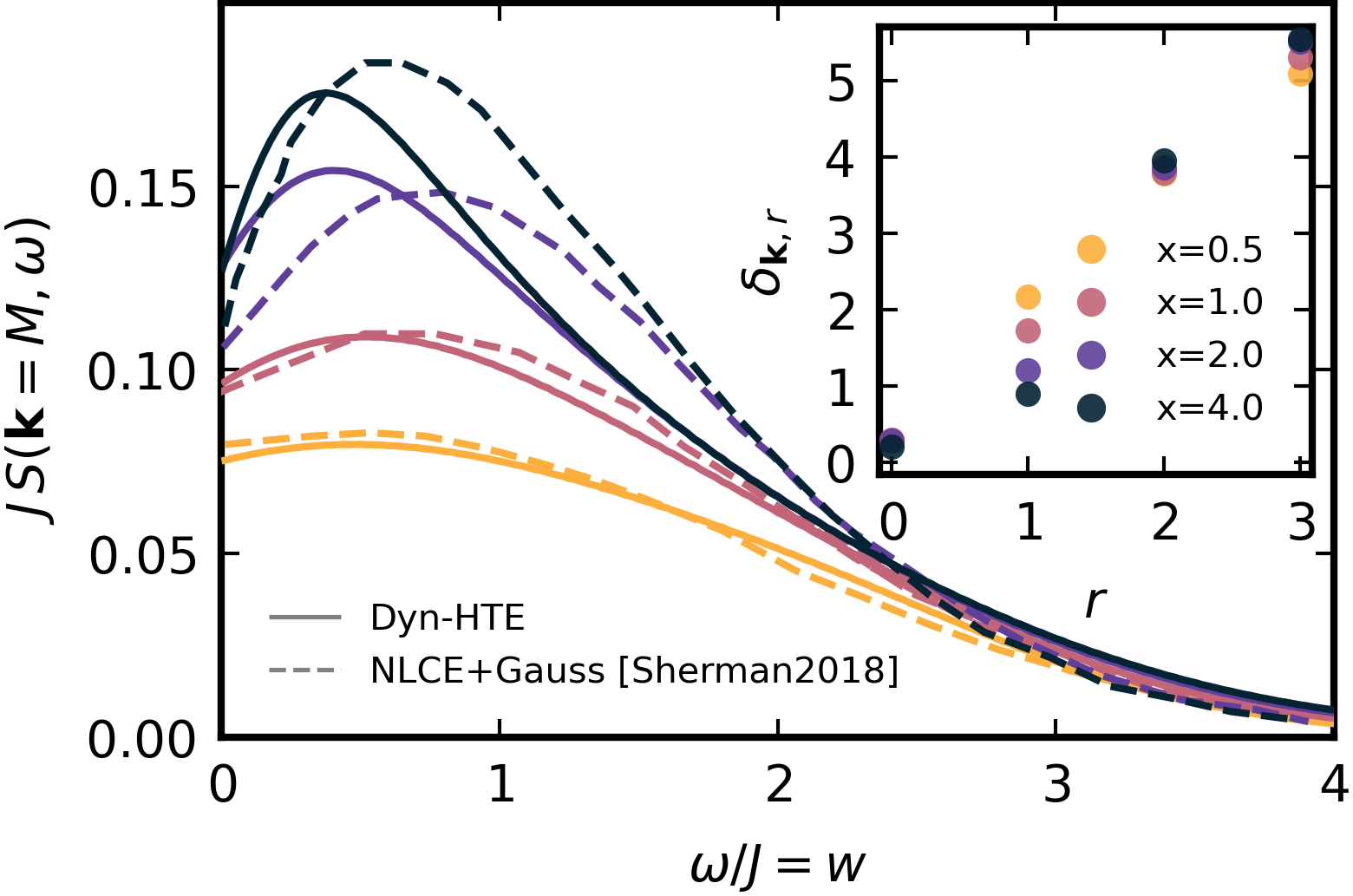}
\hspace{-2mm}
\includegraphics[scale=0.094]{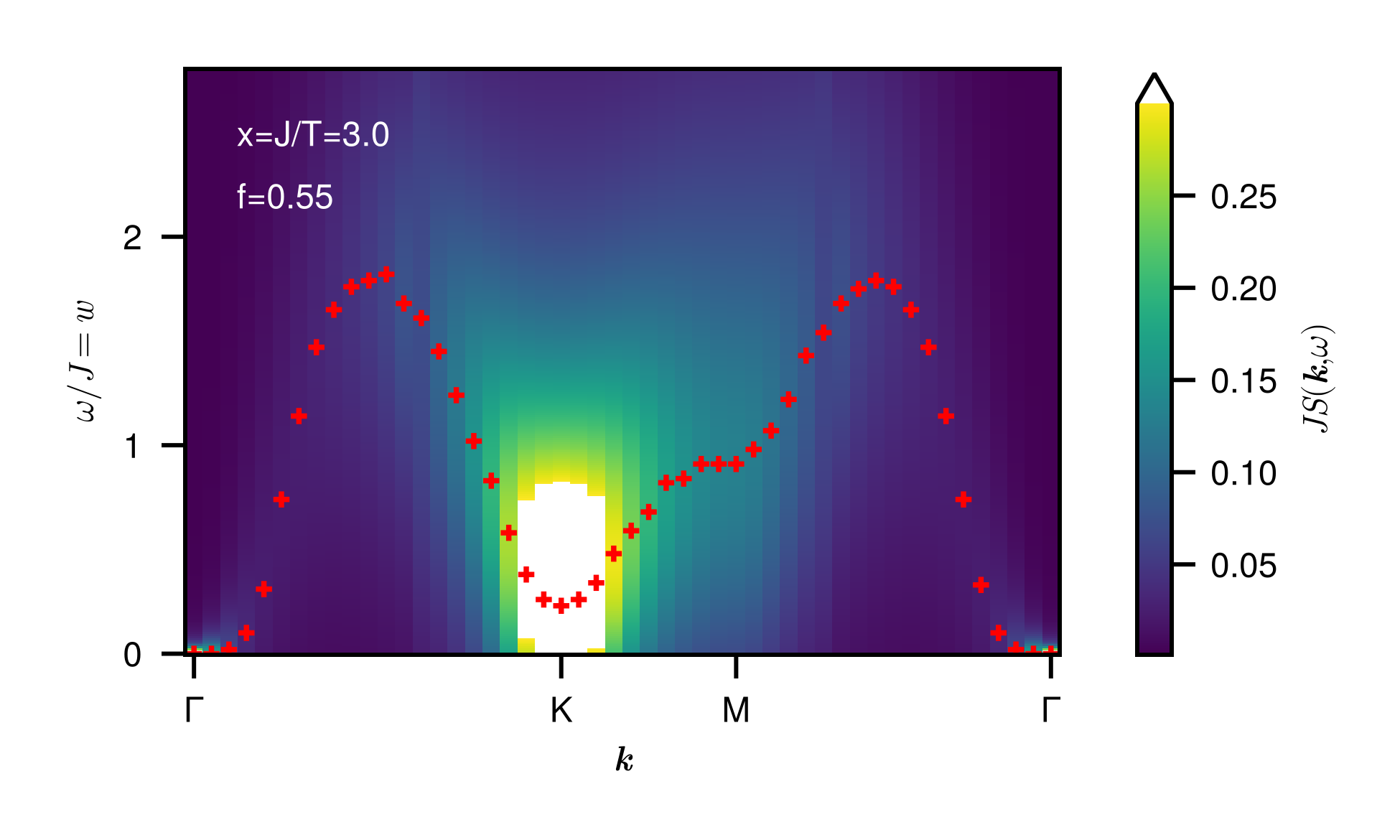}
\hspace{-5mm}
\par\end{centering}
\centering{}\caption{\label{fig:SquareLattice_KagomeLattice_JSkw} 
DSF for $S=1/2$ nearest-neighbor Heisenberg AFMs in 2D from Dyn-HTE calculated with $r_{max}=3$. 
Left: Square lattice model at $x=J/T=2$ ($f=0.7$). Crosses denote the energy of the paramagnon mode measured experimentally at $x\simeq 11$ in Ref.~\cite{dalla_piazza_fractional_2015}.
Middle: Kagome lattice model, linecuts at momentum $M$ in the extended Brillouin zone from Dyn-HTE with $r_{max}=3$ and $f=0.6$ (lines) and the NLCE results (dashed) from Ref.~\cite{sherman_structure_2018} which use only $r=0,1$ (Gaussian approximation). Inset: Continued fraction parameters $\delta_r$ for the four different temperatures $x=J/T$ from Dyn-HTE.
Right: Triangular lattice model at $x=3$ ($f=0.55$). The red crosses denote the maximum obtained from the Dyn-HTE data at fixed momentum. 
}
\end{figure*}

\emph{Square-lattice Heisenberg $S=1/2$ AFM:} This model is unfrustrated and known to exhibit long-range Néel order at $T=0$. At finite $T$, the spin correlation length becomes finite and the characteristic magnon excitations acquire a lifetime-broadening. The DSF of this model has been studied experimentally in a material realization \cite{dalla_piazza_fractional_2015} and numerically via quantum Monte Carlo \cite{shao_nearly_2017}. In Fig.~\ref{fig:SquareLattice_KagomeLattice_JSkw}(left) we present the DSF from Dyn-HTE at $x=J/T=2.0$ where the paramagnons emergent from low-$\omega$ at $\mathbf{k}=(\pi,\pi)$ are already clearly discernible. The crosses reproduce the energy of the paramagnon mode measured experimentally at $x\simeq 11$ \cite{dalla_piazza_fractional_2015}. Despite the difference in $T$, the agreement is reasonable. Another noteworthy feature of the Dyn-HTE data also observed in experiment is the suppression of the DSF at $(\pi,0)$ as compared to $(\pi/2,\pi/2)$. However, at temperatures accessible to Dyn-HTE, we fail too see the anomalous non-Gaussian DSF lineshape at $(\pi,0)$ which was proposed as a signature of spinon deconfinement \cite{dalla_piazza_fractional_2015}.

\emph{Kagome lattice Heisenberg $S=1/2$ AFM:} This frustrated nearest-neighbor model is defined by lattice vectors $\mathbf{e}_{1,2}=(1,\pm\sqrt{3})$ and a triatomic basis at the origin and at $(\pm1,-\sqrt{3})/2$. The nature of the ground state remains under debate with certain evidence for either a gapless $U(1)$ Dirac QSL \cite{ran_projected-wave-function_2007,iqbal_gapless_2013,he_signatures_2017} or a gapped $\mathbb{Z}_2$ QSL \cite{yan_spin_2011,depenbrock_nature_2012}. The DSF at $T=0$ obtained from DMRG suffers from severe finite-size effects \cite{zhu_entanglement_2018} especially at small $|\omega|$. For $x=J/T \leq 4$ the DSF has been calculated with NLCE \cite{sherman_structure_2018} and the Gaussian approximation which constructs $A_\mathbf{k}(\omega)$ from moments $r=0,1$. We focus on the high-symmetry momentum $M=(0,2\pi/\sqrt{3})$ and show the DSF from Dyn-HTE using four moments in Fig.~\ref{fig:SquareLattice_KagomeLattice_JSkw}(middle panel, full lines). Padé approximants of the $u$-series ($f=0.6$) allow us to reach $x=4$. For $x\leq 1$ the Dyn-HTE is close to NLCE results (dashed lines) but for larger $x$ the line-shapes deviate significantly. At the $K$-point, the DSF (not shown) closely resembles a slightly scaled up version of the data at the $M$-point.         

\emph{Triangular lattice Heisenberg $S=1/2$ AFM:} We supplement the linecuts shown in Fig.~\ref{fig:Triangular} with a plot of the DSF at $x=3$ along a BZ path, see Fig.~\ref{fig:SquareLattice_KagomeLattice_JSkw}(right panel). For each momentum, we denote the frequency with maximum DSF by a red cross. The shallow dip characterizing the roton-like excitation at $\mathbf{k}=M$ can already be inferred. 

\begin{figure}[b]
    \centering
    \vspace{-1mm}
    \includegraphics[width=0.85\linewidth]{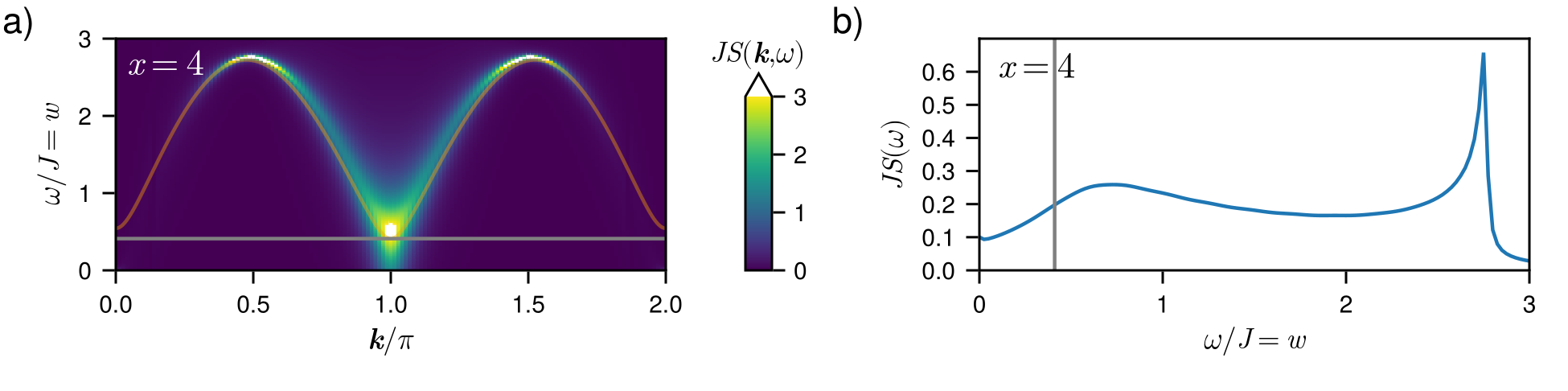}
    \vspace{-5mm}
    \caption{$S=1$ AFM Heisenberg chain. (a) DSF for  $x = J/T = 4$ from Dyn-HTE with $r_{max}= 3$ and $f=0.48$. The red line shows the maxima of the DSF at $T=0$ obtained from DMRG \cite{white_spectral_2008}. (b) On-site DSF $S(\omega)=1/(2\pi) \int_{-\pi}^\pi \mathrm{d}\mathbf{k} \, S(\mathbf{k},\omega)$ obtained from the data in panel (a). In both panels, the gray lines indicate the Haldane gap of $ 0.41049(2)J$ \cite{golinelli1994finite}.
    }
    \label{fig:S1_chain_DSF}
\end{figure}

\emph{Heisenberg $S=1$ AFM chain:} This model serves as a benchmark for $S=1$ and features a Haldane gap of $ \simeq0.41J$ \cite{golinelli1994finite}. In Fig.~\ref{fig:S1_chain_DSF}(a) we show the DSF from Dyn-HTE at $x=4$ for which the maximum of the dispersion agrees already closely to a ground state error-controlled DMRG result \cite{white_spectral_2008} (red line). The temperature-induced faint intraband magnon scattering signal \cite{beckerFinitetemperatureDynamicsThermal2017} is not resolved, likely due the low number of moments.
A practical insight for postprocessing of Dyn-HTE data is gained from the attempt to compute the real-space on-site structure factor. From ground-state simulations \cite{white_spectral_2008} it is expected to feature a double-peak structure that will be softened by temperature. In principle, one could base its calculation on the onsite moments, $m_{{ii},2r}$ obtained from the local Matsubara correlator. However, the result is more robust and closer to the expected shape if instead one takes the $\mathbf{k}$-integral over the $\mathbf{k}$-resolved DSF of panel (a), this is shown in Fig.~\ref{fig:S1_chain_DSF}(b). The reason is that single-peak shapes [as in $S(k,\omega)$] are more reliably obtained from a limited number of frequency moments $m_{\mathbf{k},2r}$ than more complicated shapes.

\end{document}